\documentclass[aip,numerical,pop,preprint,amsmath,amssymb,letterpaper]{revtex4-2}

\usepackage{graphicx}
\usepackage{dcolumn}
\usepackage{bm}
\usepackage{xcolor}
\usepackage{hyperref}
\hypersetup{colorlinks=true,citecolor=blue,linkcolor=black,urlcolor=blue}
\usepackage[super]{nth}
\usepackage{indentfirst}
\definecolor{orcidlogocol}{HTML}{A6CE39}
\usepackage{svg}

\newcommand{\vek}[1]{\boldsymbol{#1}}
\newcommand{\uvek}[1]{\hat{\boldsymbol{#1}}}

\newcommand{\bnorm}{\uvek{b}}

\newcommand{\exb}{\boldsymbol{E}\times\boldsymbol{B}}
\newcommand{\gav}[1]{\overline{#1}}
\newcommand{\Ah}{A_\parallel^{(h)}}
\newcommand{\As}{A_\parallel^{(s)}}

\newcommand{\bstar}{\boldsymbol{B}^\ast}

\newcommand{\musec}{\mu\mathrm{s}}
\newcommand{\orcid}[1]{\href{https://orcid.org/#1}{\textcolor{orcidlogocol}{\includegraphics[height=11pt]{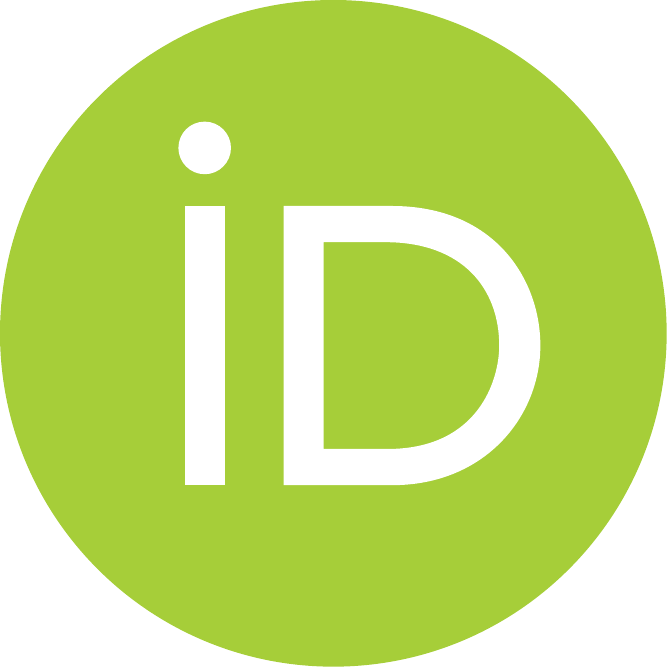}}}}

\begin{document}


\title{Electromagnetic total-f algorithm for gyrokinetic particle-in-cell simulations of boundary plasma in XGC}
\author{Robert Hager \orcid{0000-0002-4624-3150}}
 \email{rhager@pppl.gov}
\affiliation{%
Princeton Plasma Physics Laboratory\\
P.O. Box 451, Princeton, NJ 08543, USA
}%
\author{S. Ku \orcid{0000-0002-9964-1208}}%
\affiliation{%
Princeton Plasma Physics Laboratory\\
P.O. Box 451, Princeton, NJ 08543, USA
}
\author{A. Y. Sharma \orcid{0000-0002-7946-7425}}%
\affiliation{%
Princeton Plasma Physics Laboratory\\
P.O. Box 451, Princeton, NJ 08543, USA
}
\author{C. S. Chang \orcid{0000-0002-3346-5731}}%
\affiliation{%
Princeton Plasma Physics Laboratory\\
P.O. Box 451, Princeton, NJ 08543, USA
}
\author{R. M. Churchill \orcid{0000-0001-5711-746X}}%
\affiliation{%
Princeton Plasma Physics Laboratory\\
P.O. Box 451, Princeton, NJ 08543, USA
}
\author{the XGC Team}%
\affiliation{%
Princeton Plasma Physics Laboratory\\
P.O. Box 451, Princeton, NJ 08543, USA
}


\date{\today}


\begin{abstract}
The simplified $\delta f$ mixed-variable/pull-back electromagnetic simulation algorithm implemented in XGC for core plasma simulations by M. Cole et al. [Phys. Plasmas 28, 034501 (2021)] has been generalized to a total-f electromagnetic algorithm that can include, for the first time, the boundary plasma in diverted magnetic geometry with neutral particle recycling, turbulence and neoclassical physics.
The $\delta f$  mixed-variable/pull-back electromagnetic is based on the pioneering work by Kleiber and Mischenko et al. [R. Kleiber et al., Phys.
Plasmas 23, 032501 (2016); A. Mishchenko et al., Comput. Phys. Commun. 238, 194 (2019)].
An electromagnetic demonstration simulation is performed in a DIII-D-like, H-mode boundary plasma, including a corresponding comparative electrostatic simulation,  which confirms that the electromagnetic simulation is necessary for a higher fidelity understanding of the electron particle and heat transport even at the low-$\beta$ pedestal foot in the vicinity of the magnetic separatrix.
\end{abstract}


\maketitle

\section{Introduction}\label{sec:introduction}

XGC is a gyrokinetic total-$f$ particle-in-cell (PIC) code for the calculation of neoclassical physics, turbulence physics, and neutral particle dynamics together in tokamaks and stellarators.
XGC specializes in edge and scrape-off-layer (SOL) physics, but its computational domain is the whole volume enclosed by the inner wall to encompass possible non-local core-pedestal-SOL interaction.
This includes the magnetic axis, the separatrix and the scrape-off layer in contact with the divertor plates.
Until now, most of the total-$f$ physics studies have been performed with the electrostatic version of XGC due to the instabilities associated with the so called ``cancellation problem'' (see Refs. \onlinecite{cummings_phd_1995} pp. 176-177, \onlinecite{Chen_2003}, and \onlinecite{mishchenko_2004}).
A comprehensive summary of the edge-physics specific features of electrostatic XGC can be found in Ref.~\onlinecite{ku_2018}, and a description of the stellarator capability in Refs. \onlinecite{moritaka_2019,cole_2019_2}.
XGC has been extensively benchmarked against other codes, e.g., the drift-kinetic code NEO for neoclassical and impurity transport \cite{hager2016_1,hager_2019_2,Dominski2019}, and the gyrokinetic codes GENE \cite{merlo_2018}, GTC \cite{holod_2013}, GEM \cite{hager_2017}, ORB5 \cite{Cole2021}, and EUTERPE \cite{cole_2019_2} for nonlinear turbulence.

Recently, two reduced $\delta f$ electromagnetic algorithms for the core plasma region (that omit the neoclassical driver on the right-hand-side of the gyrokinetic Boltzmann equation) that are designed to mitigate the ``cancellation issue'' have been installed in XGC.
One is a fully implicit time-integration algorithm, as demonstrated by B. Sturdevant {\it et al.} in Ref.~\onlinecite{Sturdevant2021}.
The other is an explicit time-integration algorithm based on the mixed-variable/pullback formulation as shown by M. Cole {\it et al.} in Ref. \onlinecite{Cole2021}.

The purpose of this paper is to report the generalization of XGC's explicit electromagnetic algorithm from the reduced $\delta f$ algorithm into a total-$f$ electromagnetic algorithm, which allows for electromagnetic boundary plasma simulation for the first time.
This new version embraces all the edge capabilities of the previous XGC electrostatic version including the magnetic separatrix, scrape-off layer plasma, realistic divertor geometry, and neutral particle Monte-Carlo recycling and transport.

This paper is organized as follows: Section \ref{sec:algorithm} contains the major contents of the present paper, laying out details of how the reduced $\delta f$ mixed-variable/pull-back electromagnetic algorithm is upgraded to total-$f$ encompassing neoclassical, turbulence and neutral particle dynamics, together with sources, sinks and logical sheath, in contact with the material wall.
Section \ref{sec:demonstration} presents an example electromagnetic simulation in a DIII-D-like H-mode plasma, followed by conclusions and discussion in Sec. \ref{sec:summary}.


\section{Description of XGC's Electromagnetic Total-f Algorithm} \label{sec:algorithm}

The total-$f$ gyrokinetic code XGC solves the five-di\-mensional (5D) gyrokinetic Boltzmann equation
\begin{equation}\label{eq:xgc_boltzmann}
    \frac{\mathrm{d} f}{\mathrm{d}t} = S(f),
\end{equation}
using the nonlinear marker particle equations of motion by Kleiber \textit{et al.} in Ref. \onlinecite{Kleiber_2016}.
The equations of motion to be described in the present paper are based on the $\delta f$ mixed-variable formulation \cite{mishchenko_2014_1,Kleiber_2016} and the pullback-transformation method \cite{mishchenko_2014_2,Mishchenko2019}, and were developed to avoid the notorious ``cancellation problem'' in Amp\`{e}re's law \cite{Chen_2003} while allowing for the use of an explicit time integrator and large time steps compared to the electron skin time.
In Eq. \eqref{eq:xgc_boltzmann}, the term $S(f)$ represents generalized source and dissipation terms, e.g., Fokker-Planck collisions \cite{yoon_2014,hager2016_2}, heat and torque input \cite{ku_2018,hager_2019_1}, and atomic interactions with neutral particles \cite{stotler_2017}.

In this section, we describe our generalization of the electromagnetic $\delta f$ mixed-variable equations of motion to XGC's total-$f$ formalism.
We provide the exact form of the equations of motion, the gyrokinetic Poisson equation and Amp\`{e}re's law.
We also describe XGC's representation of marker particles and the background distribution function, the weight-update equation, the total-$f$ pullback transformation, and the space-time discretization used in this electromagnetic version of XGC.

\subsection{Equations of motion}\label{subsec:eqs_of_motion}

The mixed-variable formulation of the gyrokinetic equations of motion \cite{mishchenko_2014_1} splits the parallel component of the perturbed magnetic vector potential $A_\parallel = \bnorm \cdot \vek{A}$ (with $\bnorm = \vek{B_0}/B_0$) into two parts, called the Hamiltonian ($\Ah$) and the symplectic ($\As$) parts:
\begin{equation}\label{eq:apara_mixed_def}
    A_\parallel = \Ah+\As.
\end{equation}
How $A_\parallel$ is split into these two parts is determined by the choice of the constraint for $\partial \As/\partial t$ (see Sec. \ref{subsec:pullback}).
The introduction of the symplectic vector potential $\As$ is meant for mitigation of the cancellation problem.

The parallel velocity coordinate is defined as $u_\parallel = v_\parallel + (q_s/m_s) \gav{\Ah}$, where $v_\parallel = \bnorm \cdot \vek{v}$  is the velocity along the equilibrium magnetic field, $q_s$ and $m_s$ are the charge and mass of the particle of species $s$, and an overline indicates the gyroaverage.
The full set of 5D phase-space coordinates is $(\vek{R},u_\parallel,\mu_s)$, where $\vek{R}$ is the gyrocenter position in configuration space and $\mu_s = m_s v_\perp^2/(2 B_0)$ is the magnetic moment.

The Lagrangian equations of motion (under the condition $\dot{\mu}_s=0$) in these coordinates are \cite{Kleiber_2016}
\begin{align}
\dot{\vek{R}} &= \frac{D}{B_0} \left[ \frac{\bstar}{m_s}\,\frac{\partial H}{\partial u_\parallel} + \frac{\vek{F}\times\vek{B}_0}{B_0} \right], \label{eq:xgc_rdot} \\
\dot{u}_\parallel &= \frac{q_s}{m_s} \left[ D \frac{\bstar}{B_0} \cdot \vek{F} - \frac{\partial \As}{\partial t} \right], \label{eq:xgc_udot}
\end{align}
where $\vek{B}_0=\nabla \times \vek{A}_0$ is the equilibrium magnetic field, $B_0=|\vek{B}_0|$, $\bstar=\nabla \times \vek{A}^\ast$, and
\begin{align}
\vek{A}^\ast &= \vek{A}_0 + \left(\frac{m_s}{q_s} u_\parallel + \gav{\As}\right) \bnorm , \label{eq:xgc_astar}\\
H &= \frac{m_s}{2} u_\parallel^2 + \mu_s B_0 + q_s \left( \gav{\phi} - u_\parallel \gav{\Ah} \right) + \frac{q_s^2}{2 m_s} \gav{\Ah}^2, \label{eq:xgc_hamiltonian} \\
\frac{\partial H}{\partial u_\parallel} &= m u_\parallel - q_s \Ah, \\
\vek{F} &= -\frac{1}{q_s}\,\frac{\partial H}{\partial \vek{R}} = -\frac{\mu_s}{q_s} \nabla B_0 - \nabla\gav{\phi} + u_\parallel \nabla\gav{\Ah} - \frac{q_s}{m_s} \gav{\Ah} \left( \nabla \gav{\Ah} \right), \label{eq:xgc_ham_force} \\
D &= \frac{B_0}{\bnorm \cdot \bstar} = \left[ 1 + \left( \frac{m_s u_\parallel}{q_s B_0} + \frac{\gav{\As}}{B_0} \right) \bnorm \cdot \nabla \times \bnorm \right]^{-1}. \label{eq:xgc_D}
\end{align}

\subsection{Field equations}\label{subsec:field_equations}

The field equations are the gyrokinetic Poisson equation for the electrostatic potential $\phi$ and the mixed-variable Amp\`{e}re's law for the Hamiltonian component of the magnetic vector potential $\Ah$.

The gyrokinetic Poisson equation with the Pad\'{e} approximation is
\begin{equation}\label{eq:xgc_gk_poisson}
 \nabla\cdot \frac{n_0 m_i}{q_i B_0^2} \left( 1+\frac{T_i}{T_e} \right) \nabla_\perp \phi = -\left( 1+ \nabla \cdot \rho_i^2 \nabla_\perp \right) \left(\overline{n}_i - n_e \right).
\end{equation}
Here, $n_0$ and $T_{i/e}$ are the background density and ion/electron temperature, and $\rho_i=(m_i k_B T_i)^{1/2}/(q_i B_0)$ is the ion gyroradius.
In the long-wavelength approximation, the polarization density term proportional to $T_i/T_e$ on the left-hand side and the term $\rho_i^2$ on the right-hand side of Eq. \eqref{eq:xgc_gk_poisson} are neglected.

The mixed-variable Amp\`{e}re's law is given by
\begin{equation}\label{eq:xgc_ampere}
  -\nabla \cdot \nabla_\perp \Ah + \Ah \sum_{s=i,e} \frac{\mu_0 n_0 q_s^2}{m_s}
       = \mu_0 \left( \gav{\delta j_{\parallel,i}} + \delta j_{\parallel,e} \right) + \nabla \cdot \nabla_\perp^2 \As,
\end{equation}
where $\mu_0$ is the permeability of free space, and the parallel current densities are given by the first $u_\parallel$ moments of the species distribution functions.

\subsection{Mixed PIC-Eulerian Representation}\label{subsec:mixed_rep}

Equation \eqref{eq:xgc_boltzmann} is discretized with a mixture of Lagrangian PIC and Eulerian grid methods \cite{sku_2016,hager2016_2}.
The left-hand side is evaluated using a PIC method, whereas the right-hand side is evaluated on a 5D phase space mesh (see Sec. \ref{subsec:space-time}).

The distribution function $f$ is split into three parts: i) an analytic background $f_a$; ii) an Eulerian contribution $f_g$ on a 5D phase-space mesh; and iii) a contribution represented by marker particles, $f_p$, i.e.,
\begin{equation}\label{eq:xgc_f_splitting}
    f = f_a + f_g + f_p.
\end{equation}
For clarity of the discussion, we use the combined analytic and Eulerian contributions, $f_0=f_a+f_g$, to form the background for the fast space-time dynamics of the particle part $f_p$.
In other words, the background $f_0$
varies (see Sec. \ref{subsec:f_analytic_update}) on a slower timescale than $f_p$, which contains the turbulence fluctuations and particle orbit motion.

The time evolution equation of the particle distribution function $f_p$ can then be written as
\begin{equation}\label{eq:xgc_dfpdt}
\frac{\mathrm{d} f_p}{\mathrm{d}t} = -\frac{\mathrm{d} f_0}{\mathrm{d} t} + S(f),
\end{equation}
which is equivalent to Eq. \eqref{eq:xgc_boltzmann}.
Here, $f_p$ describes marker particles that move according to equations \eqref{eq:xgc_rdot}, \eqref{eq:xgc_udot}, \eqref{eq:xgc_gk_poisson}, and \eqref{eq:xgc_ampere}, and are represented as $f_p = w_0 w_1 g$ with the marker particle distribution function (or inverse phase space volume) $g$.
The time-invariant ``full-$f$'' marker weight $w_0$ is obtained from importance sampling of the initial plasma distribution function $\left.f\right|_{t=0} = f_0\left.\right|_{t=0}$ such that $w_0 = (f_0/g)\left.\right|_{t=0}$.
Here, the notation $f\left.\right|_{t=t_0}$ implies the value of $f$ at the phase space position at time $t=t_0$ on the trajectory of markers. 
Since $\mathrm{d}g/\mathrm{d}t=0$ along the particle trajectories, the time evolution equation of the varying (``delta-$f$'') marker weights $w_1$ is
\begin{equation}\label{eq:xgc_dwdt}
    \frac{\mathrm{d}w_1}{\mathrm{d}t} = \frac{1}{f_0\left.\right|_{t=0}} \left( -\frac{\mathrm{d} f_0}{\mathrm{d} t} + S(f) \right).
\end{equation}

An operator splitting is employed to separate the weight evolution from particle motion and sources in Eq. \eqref{eq:xgc_dwdt}.
The operator splitting is valid since these two operations are independent and commute.
The source contribution becomes
\begin{equation}\label{eq:xgc_dwdt_source}
    \left. \Delta w_1 \right|_{\mathrm{source}} = S(f) \Delta t,
\end{equation}
with the time step $\Delta t$.
In XGC's total-$f$ method, the contribution from particle motion is evaluated by direct integration,
\begin{equation}\label{eq:xgc_dwdt_motion}
    \left. \Delta w_1 \right|_{\mathrm{motion}} = \frac{f_0\left.\right|_{t=t_0}-f_0\left.\right|_{t=t_0+\Delta t}}{f_0\left.\right|_{t=0}}.
\end{equation}

\subsection{Pullback transformation}\label{subsec:pullback}

In order to close the equations of motion and the field equations \eqref{eq:xgc_rdot}, \eqref{eq:xgc_udot}, \eqref{eq:xgc_gk_poisson}, and \eqref{eq:xgc_ampere}, a constraint must be chosen for $\partial \As/\partial t$.
The scheme utilizes $\As$ to maximize the numerical stability of the electromagnetic gyrokinetic equation.
Two choices are implemented in XGC:
\begin{align}
    \mathrm{i)}&\quad \frac{\partial \As}{\partial t} = 0, \label{eq:xgc_pb3}\\
    \mathrm{ii)}&\quad \frac{\partial \As}{\partial t} = -\bnorm \cdot \nabla \phi. \label{eq:xgc_pb4}
\end{align}
%
Note that Eq. \eqref{eq:xgc_pb4} is reminiscent of the ideal MHD limit of $E_\parallel=0$.

In order to limit the growth of the ``cancellation-causing'' Hamiltonian part and keep $\Ah \ll \As$, $\Ah$ is reset to zero after every time step while keeping $\vek{A}=\As+\Ah$ constant. 
We refer the readers to Refs. \onlinecite{Kleiber_2016,Mishchenko2019} for a detailed description of the scheme in the $\delta f$ framework.
Under this coordinate ``pullback'' transformation
\begin{equation}
    u_\parallel^{(old)} \rightarrow u_\parallel^{(new)}=u_\parallel^{(old)}-\frac{q_s}{m_s}\gav{\Ah},
\end{equation}
and the total distribution function remains invariant, i.e., $\tilde{f}(u_\parallel^{(new)}) = f(u_\parallel^{(old)})$.
This constraint defines the corresponding transformation of the marker particle weights in a PIC scheme.
Since the background $f_0$ is left unchanged in the pullback transformation process while the parallel speed undergoes a coordinate transformation, the corresponding change of the marker particle weight in the present total-$f$ algorithm is [see Eq.~\eqref{eq:xgc_dwdt_motion}]
\begin{equation}
    \left.\Delta w_1 \right|_{\mathrm{pullback}} = \frac{f_0(u_\parallel^{old})-f_0(u_\parallel^{new})}{\left. f_0\right|_{t=0}}.
\end{equation}

\subsection{Evolution of the background distribution function}\label{subsec:f_analytic_update}


The total-$f$ code XGC employs not only the time-constant full-f weight $w_0$, but also the time-evolving ``delta-f'' weight $w_1$ that contains information on the evolution of background plasma profile and, at the same time, suffers from particle noise growth.
In the total-$f$ method in XGC, the growth of the marker weights $w_1$ is transferred to the background distribution function $f_0$, a technique first introduced in Ref. \onlinecite{sku_2016}, to allow for the evolution of the $f_0$ and the mitigation of the growing-weight issue.

Two methods are implemented and can be used alone or in combination.
Both methods transfer a small fraction of the marker weights to the background distribution function in each time step, i.e.,
\begin{equation}
    f_0^{(new)} = f_0^{(old)} + \alpha \mathcal{P}(f_p),
\end{equation}
where $\alpha \ll 1$ and $\mathcal{P}$ is a linear ``projection'' operator.
$\alpha$ needs to be small to ensure that the timescale for the $f_0$ time-evolution is slower than the turbulence eddy and particle dynamics timescale.
The operator $\mathcal{P}$ can be i) the projection on a 5D phase space grid 
as described in Ref. \onlinecite{hager2016_2} [hence time-advancing $f_g$ in Eq. \eqref{eq:xgc_f_splitting}]; or ii) a projection to a Maxwellian to modify $f_a$ with $\mathcal{P}(f_p)$ representing the density, parallel flow and kinetic energy moments of $f_p$.

While method i) has the advantage of being able to represent any non-Maxwellian distribution function, it is limited to velocities below $v_{max}$ used in our velocity grid, which can limit the role of the nonlocal higher energy tail particles in the H-mode pedestal.
Method ii), on the other hand, captures changes of the background at all velocities. 
Method ii) can be generalized by transferring the anisotropic temperature information from $f_p$ to $f_a$ or using a set of Maxwellian basis functions \cite{chen2021unsupervised}.

\subsection{Time and space discretization}\label{subsec:space-time}

Configuration space $\vek{R}$ is discretized with (approximately) magnetic field-aligned, unstructured triangle meshes in right-handed cylindrical coordinates $(R,\varphi,Z)$ \cite{adams_ku_2009,fzhang_2015}.
Similar to magnetic flux-coordinates, the field-alignment allows to resolve perturbations with low parallel wavenumber $k_\parallel=\bnorm \cdot \vek{k}$ (compared to $k_\perp$) and high toroidal mode number with relatively low toroidal resolution (see Ref. \onlinecite{Hariri2015} for comparison).

In the toroidal direction $\uvek{\varphi}$, the mesh spacing is uniform, i.e., $\Delta\varphi = 2\pi/(N_w N_\varphi)$, where $N_\varphi$ is the toroidal grid size and $N_w\geq 1$ is the wedge fraction.
$N_w$ is unity if the simulation domain is the whole torus from $\varphi=0$ to $2\pi$, and $N_w>1$, if only a fraction of the torus from $\varphi=0$ to $\varphi=2\pi/N_w$ is simulated with periodic boundary conditions in the toroidal direction.

Simulating only a wedge of the full torus makes XGC simulations more economical in terms of use of computing time, but thins the captured toroidal mode spectrum.
The properties of the wedge approximation have been studied in XGC in Ref.~\onlinecite{Kim2017}.

\begin{figure}
 \centering
 \includegraphics{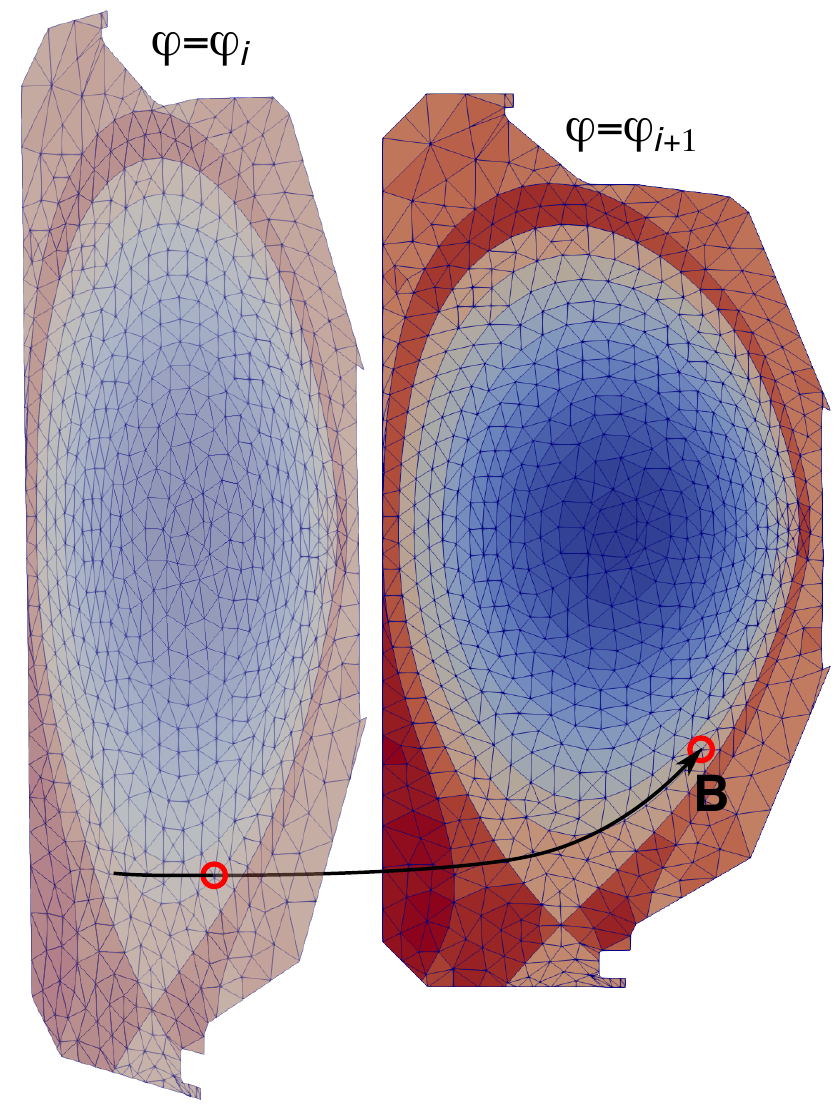}
 \caption{Schematic illustration of the approximately field-aligned structure of XGC meshes.}
 \label{fig:pic_cell}
\end{figure}

At each toroidal grid point $\varphi_i$, the radial-poloidal plane is discretized with unstructured triangle meshes and linear interpolation between mesh vertices in the $R$ and $Z$ directions.
The triangle meshes are approximately aligned to the equilibrium magnetic field in the sense that a field-line starting on a mesh vertex at $\varphi=\varphi_i$ intersects the plane $\varphi=\varphi_{i+1}$ at or very close to another vertex \cite{adams_ku_2009,fzhang_2015} (Fig. \ref{fig:pic_cell}).
%
Particle scatter (charge deposition for the field solver) and gather (interpolation of fields at the particle positions) operations between toroidal grid points are not executed along the toroidal direction, but rather along the magnetic field, i.e., aligned with the slow variation of low-$k_\parallel$ perturbations.
Interpolation along the magnetic field is linear.
For a detailed description, we refer the readers to Ref. \onlinecite{moritaka_2019}.
%

In order to resolve the ion gyroradius-scale microturbulence, XGC meshes have high resolution [$\mathcal{O}(\rho_i)$] within the radial-poloidal plane.
The (full-torus) toroidal resolution can be relatively low, typically $N_\varphi\sim$32-64 on currently available high-performance computing (HPC) systems for $N_w=1$ and higher for greater $N_w$.
As long as the parallel wavenumber $k_\parallel$ is small and the poloidal resolution is high enough (e.g., Ref. \onlinecite{Hariri2015}), the field-alignment allows for resolving perturbations with toroidal mode numbers $n$ greater than the number of toroidal grid points.

%
%

Velocity space is discretized with uniform rectangular grids in the parallel and perpendicular (to $\bnorm$) velocities $(v_\parallel,v_\perp)$.
The velocity grids at each vertex of the configuration space mesh $\vek{R}_i$ are normalized to the local thermal speed $v_{th,s}(\vek{R}_i) = [k_B T_s(\vek{R}_i)/m_s]^{1/2}$ with a cutoff of $v_{max}=N_v v_{th,s}(\vek{R}_i)$.
Each particle species has its individual $v_{max}$ even though they share the same normalized $v_{max}/v_{th,s}(\vek{R_i})$.

The time advance of particles and fields is implemented with a hybrid of \nth{2} and \nth{4} order Runge-Kutta (RK) methods.
Each of the two RK2 stages consists of i) an RK4 particle push with fixed electric and magnetic fields according to Eqs. \eqref{eq:xgc_rdot} and \eqref{eq:xgc_udot}, ii) charge and current deposition, and iii) solving the field equations \eqref{eq:xgc_gk_poisson} and \eqref{eq:xgc_ampere}.
In step i), electrons can be advanced with a smaller time step (and a corresponding larger number of RK4 push steps).
After step iii) in the second RK2 stage, the pullback transform is executed so that the subsequent evaluation of $S(f)$ can be done with $u_\parallel=v_\parallel$.

The source terms $S(f)$, which are not described here, are evaluated after the particle-field time-advance on a 5D phase space mesh.
The required transformations of $f_p$ between Lagrangian and Eulerian representations are discussed in Refs.~\onlinecite{yoon_2014,mollen_2021}.
The nonlinear Fokker-Planck collision operator employs a backward Euler time integrator \cite{yoon_2014,hager2016_2}.
Heat and torque input, and neutral particle recycling are implemented with a simple forward Euler time integrator \cite{ku_2018,hager_2019_1,stotler_2017}.

\subsection{Treatment of the axisymmetric mode}\label{subsec:axisym_mode}

Unlike in ``conventional'' $\delta f$ gyrokinetic codes specialized for core plasma simulations, treatment of the axisymmetric components of $\phi$ and $A_\parallel$ is delicate in a total-$f$ gyrokinetic algorithm because i) neoclassical physics including the bootstrap current are simulated together; and ii) the plasma could be susceptible to violent global MHD instabilities, such as vertical displacement events.
The effect of the neoclassical equilibrium current is usually included in the reconstruction of the equilibrium magnetic field $\vek{B}_0$.
Using the bootstrap current in Eq. \eqref{eq:xgc_ampere} would create an unphysical overlap between the gyrokinetic equations and the magneto-hydrodynamic (MHD) reconstruction of the axisymmetric background field.
Thus, the neoclassical bootstrap current is evaluated and removed from Eq. \eqref{eq:xgc_ampere}.

The violent vertical displacement instabilities fall outside of the validity of the gyrokinetic equations used in XGC.
When this happens in a particular discharge, we can remove the electromagnetic effect in the $n=0$ mode by splitting Eq. \eqref{eq:xgc_gk_poisson} into its axisymmetric (toroidal mode number $n=0$) and non-axisymmetric components, and applying XGC's electrostatic algorithm \cite{ku_2018} to the axisymmetric mode, i.e., $\As(n=0)=0$ and $\Ah(n=0)=0$.
Depending on the plasma, there could also be other violent global MHD instabilities seen in the total-$f$ XGC simulations.
An example is the sausage mode.
In this case, we need to remove the $m=0$ electromagnetic modes.
%

\subsection{Field-aligned Fourier filter}\label{subsec:fourier}

A field-aligned Fourier filter can be applied in XGC to avoid possible numerical instabilities caused by high-frequency Alfv\'{e}n waves that would normally require extremely small time steps and consume a larger amount of computing time than is available.
It is effectively a low-pass filter in the parallel wavenumber $k_\parallel$.

The filter was first described in Ref. \onlinecite{hager_2019_1}.
It operates in straight field line coordinates $(\psi_N,\varphi,\theta^\ast)$, where $\psi_N$ is the normalized poloidal magnetic flux, $\varphi$ is the toroidal angle, and
\begin{equation}\label{eq:theta_star}
  \theta^\ast = \frac{1}{q(\psi_N)} \int_0^{l_\theta} \frac{B_T}{R B_P} \mathrm{d}l_{\theta}^\prime,
  \quad q(\psi_N) = \frac{1}{2 \pi} \int_0^{l_{\theta,max}} \frac{B_T}{R B_P} \mathrm{d}l_{\theta}^\prime.
\end{equation}
Here, $l_\theta$ is the arc length along a flux-surface in the poloidal direction, $q(\psi_N)$ is the safety factor, and $0 \dots l_{\theta,max}$ corresponds to integration over one poloidal circuit in the closed field line region.
In the scrape-off-layer, the integration is from wall strike point to strike point.
The straight field-line angle $\theta^\ast$ is well defined everywhere except for the separatrix and on the magnetic axis (where it is irrelevant).
In order to make the integrand in Eq. \eqref{eq:theta_star} numerically integrable on the separatrix ($B_P=0$ on the X-point), we replace the value of $B_T/(R B_P)$ on the X-point by its mean on the two mesh vertices adjacent to the X-point.
Also, only the closed-loop part of the separatrix is used for this calculation: The separatrix legs are ignored.

This definition also works for a discrete Fourier transform (DFT) in the scrape-off layer.
But greater care must be taken in the SOL to choose appropriate boundary conditions.
One option is to use a window function to make the data periodic between the two strike points of an open field line.
Another is to only retain sine-modes in the SOL, implying that the filtered result must be zero on the wall boundary (analogous to the boundary conditions used in the Poisson and Amp\`{e}re solvers).

An upper bound on the Alfv\'{e}n frequency is introduced by removing modes in $\phi$, $\Ah$, and $\As$ with poloidal mode number $m$ and toroidal mode number $n$ that satisfy $|m-n q|>M$ ($M\in\mathbb{N}$).
Thus, the mode cutoff $M$ and the timestep $\Delta t$ can be chosen as a compromise between physics fidelity and computational cost.

The resolution or Nyquist limit of the Fourier filter is determined by two quantities: i) the largest difference of $\theta^\ast$ between two adjacent vertices on the same flux-surface, $\max(\Delta\theta^\ast)$; and ii) the toroidal grid size $N_\varphi$ via the condition $|m/q-n| \leq N_\varphi/2$, where $|m/q-n|$ is the number of mode periods along the magnetic field over one toroidal circuit.
Thus, the maximal resolved toroidal mode number is $n_{max}(\psi_N) = \pi/[q(\psi_N) \max(\Delta \theta^\ast)(\psi_N)]$ with the constraint on the parallel wavenumber $|m/q-n| \leq N_\varphi/2$.


\section{Electromagnetic Gyrokinetic Simulation in DIII-D Geometry}\label{sec:demonstration}

As a demonstration of XGC's electromagnetic total-$f$ capability in realistic geometry, we simulate a low magnetic-field, DIII-D-like H-mode plasma (based on discharge \#132007) in which we modified the pedestal plasma profile somewhat to minimize the large-scale transient GAM (geodesic acoustic mode) oscillations that brings the pedestal plasma to be closer to a gyrokinetic equilibrium with the given magnetic field configuration.
The discharge we used had carbon temperature in lieu of the main ion temperature.  Our choice for the pedestal plasma profile modification does not guarantee that the present simulation represents the realistic experimental condition.

\subsection{Simulation setup}\label{subsec:sim_setup}

\begin{figure}
 \centering
 \includegraphics[width=0.95\textwidth]{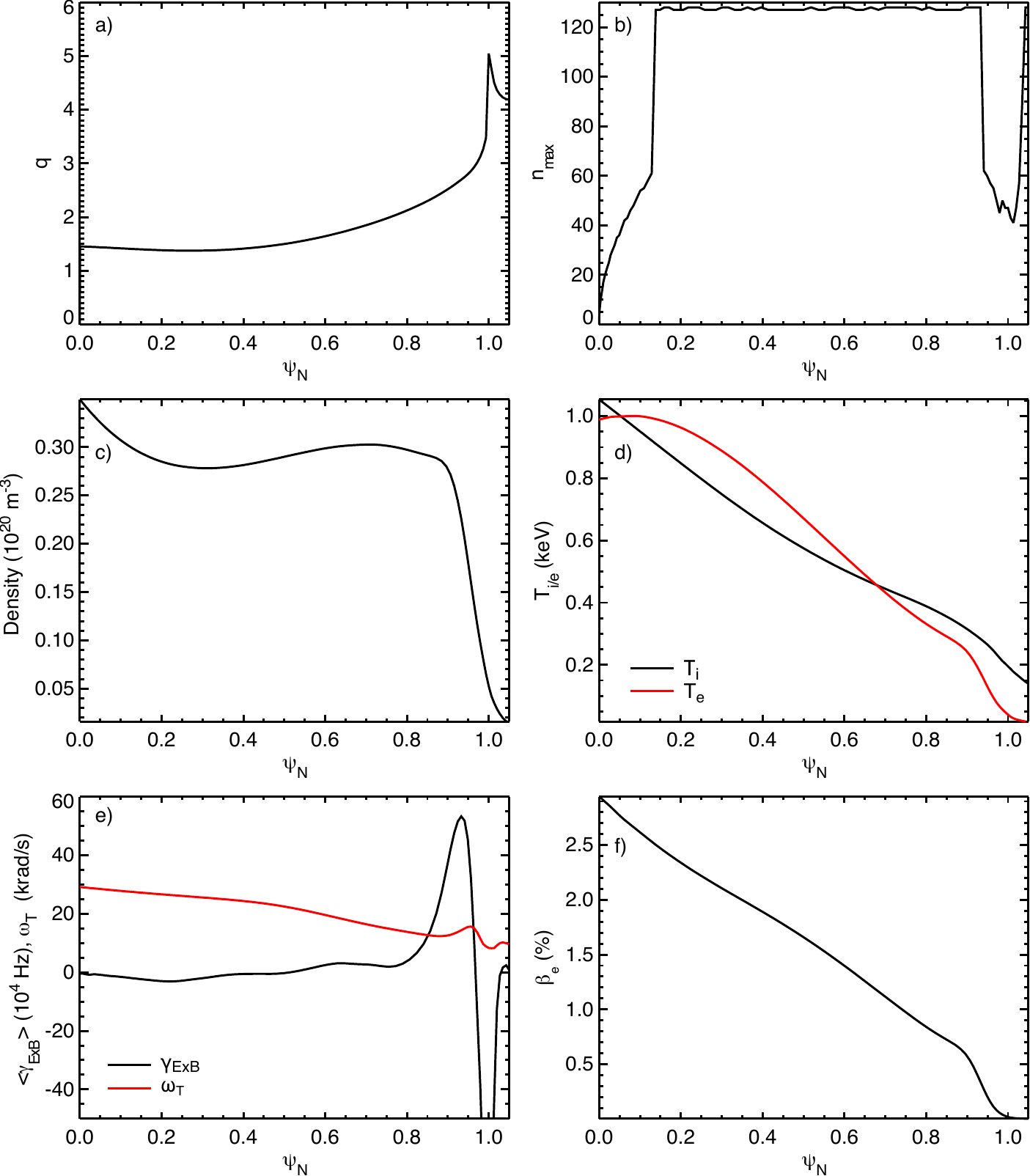}
 \caption{Initial conditions for XGC turbulence simulations at $t=t_{QS}=81 \,\musec$:
          (a) safety factor $q$; (Note that $q(\psi_N=1)$ is not infinite because of the approximation explained in Sec. \ref{subsec:fourier}.)
          (b) Nyquist limit $n_{max}$ for the toroidal mode number of the XGC mesh.
          (c) density;
          (d) ion and electron temperature;
          (e) $\exb$ shearing rate $\gamma_{\exb}$ and toroidal rotation frequency $\omega_T$. $\omega_{T}>0$ implies counter-clockwise (co-current) rotation when looking from above;
          (f) $\beta_e=2\mu_0 p_e(\psi_N)/B_0^2$.
}
 \label{fig:initial_profiles}
\end{figure}

We run one total-$f$ simulation in the electrostatic limit \cite{ku_2018}, and one with the new electromagnetic method for comparison.
The magnetic field is $B_{0}=0.687$ T on the magnetic axis and is in the clockwise direction looking down on the tokamak from above (counter-current).
The ion magnetic drift is towards the single X-point.
The major radius of the magnetic axis is $R_0=1.779$ m, the minor radius on the separatrix is $a=0.563$ m, so that the inverse aspect ratio is $\epsilon=a/R_0=0.317$.
The relatively small value of $a/\rho_i(\psi_N=0.95)=145$ in this plasma considerably reduces the computational cost that comes with electromagnetic gyrokinetic simulations.
The safety factor at $\psi_N=0.95$ is $q_{95}=2.81$ [Fig. \ref{fig:initial_profiles} (a)].

The numerical resolution in the toroidal direction is $N_\varphi=32$ and the toroidal wedge number is $N_w=1$ for the present electromagnetic simulation, and $N_\varphi=16$ with $N_w=2$ for the present electrostatic simulation.
The mesh on the radial-poloidal planes at the toroidal angle $\varphi=\varphi_i$ has a total of 79,662 vertices with a radial resolution on the outer midplane of $\Delta l_R/\rho_i \approx 1$ in the core, and $\Delta l_R/\rho_i\approx 0.5$ in the steep-gradient region between $0.9 \leq \psi_N \leq 1.02$.
The poloidal resolution $\Delta l_\theta/\rho_i$ on the outer midplane varies between 1 and 2.
Due to the field-line following nature of XGC in the cylindrical coordinate system, $\Delta l_\theta$ can be much smaller on the inboard than the outboard side.

For the present mesh, the effective toroidal resolution as described in Sec. \ref{subsec:fourier} is shown in Fig. \ref{fig:initial_profiles} (b).
$n_{max}=128$ between $0.15\leq\psi_N\leq 0.93$, and 40 to 60 between $0.94\leq \psi_N \leq 1.03$ (with a minimum of 41 at $\psi_N\sim 1.01)$.
$n_{max}$ is lower across the magnetic separatrix surface, where the safety factor $q$ is high, to avoid excessively fine poloidal mesh resolution in XGC's approximately field line following mesh arrangement.
In the present simulations, we retain toroidal mode numbers $5< n \leq \min(48,n_{max}/2)$ with $|m-nq|<6$, and the electrostatic axisymmetric mode.
The low toroidal mode numbers between 1 and 5 are excluded to avoid possibly violent global MHD instabilities (e.g. sawtooth oscillations and sausage type modes) that are not of central interest for this initial demonstration.

The electromagnetic simulation time step for the ions, which is equal to the Poisson and Amp\`{e}re's law solver time step, is $\Delta t=5.8\cdot 10^{-2}\tau_A(\psi_N=0.95)=4.04\cdot 10^{-8}$ s with $\tau_A(\psi_N)=R_0 (\mu_0 n(\psi_N) m_i)^{1/2}/B_0$.
In the electrostatic reference simulation, $\Delta t$ is four times larger.
The (subcycled) electron time step for this study is $\Delta t/4$ in the electromagnetic, and $\Delta t/20$ in the electrostatic simulation.
In order to reduce the required computational resources in this demonstration, we use artificially heavy electrons with an electron-proton mass ratio of $m_e/m_p=5\cdot 10^{-3}$.

A 2 MW heat source is applied in the core plasma.
The heating power is evenly split between ions and electrons and spread over the region $0\leq \psi_N \leq 0.9$ with a shape function $\Lambda(\psi_N)$ (see Ref. \onlinecite{hager_2019_1}) that is constant for $0.2 \leq \psi_N \leq 0.7$ and linearly tapered off in a boundary layer of width $\Delta\psi_N=0.2$.
0.5 MW of electron cooling is applied at $0.995 \leq \psi_N \leq 1.08$ with a buffer layer of $\Delta\psi_N=0.005$ in order to emulate radiative energy losses in the SOL.
A counter-current (co-$\vek{B}_T$) torque of 1.4 Nm is applied with the same shape function as the ion and electron heating.

Coulomb collisions are modeled with a nonlinear Fokker-Planck-Landau collision operator \cite{yoon_2014,hager2016_2}.
The only particle source is from neutral particle recycling in the SOL \cite{stotler_2017}.

We employ a quiet start technique that is commonly used in XGC simulations to step over the initial transient relaxation caused by the initialization of the marker particles with local shifted Maxwellians (see Ref. \onlinecite{hager_2020}).
In the first 500 time steps, corresponding to $t_{QS}=81$ $\musec$ or roughly one toroidal transit time of a 200 eV ion, the non-axisymmetric electrostatic and vector potentials are set to zero, and only the axisymmetric electrostatic potential, i.e. neoclassical physics, is solved in XGC.
In order to prevent an unphysical pedestal build-up while there is no turbulent transport, the neutral recycling rate at $t < t_{QS}$ is lowered to 80\%.
After the neoclassical quiet-start period, the non-axisymmetric components of $\phi$ and $A_\parallel$ are evaluated self-consistently, and the recycling rate is set to 99\%.
The quiet-start period used here is typically long enough for the initial transient to decay at least in the edge plasma so that turbulence can be observed on a quiescent background.

The flux-surface averaged plasma density, ion and electron temperatures, $\exb$ shearing rate $\gamma_{\exb}=\left\langle (\nabla \psi_N/|\nabla \psi_N|) \cdot \vek{v}_{\exb} \right\rangle$ and toroidal rotation frequency $\omega_T=\left\langle u_T/R \right\rangle$ (where $u_T$ is the mean toroidal flow), and $\beta_e=2\mu_0 p_e(\psi_N)/B_0^2$ at the end of the quiet start period ($t=t_{QS}$) are shown in Fig. \ref{fig:initial_profiles} (c)-(f).

The electrostatic simulation was executed on the Cori KNL system at the National Energy Research Scientific Computing Center (NERSC) using 2,048 compute nodes for approximately 45 hours ($\approx$ 90,000 node hours) with a total of 20 billion marker particles.
The electromagnetic simulation was executed on the more powerful Summit system at the Oak Ridge Leadership Computing Facility (OLCF) using 2,048 compute nodes for about 32 hours ($\approx$ 65,000 node hours) with a total of 32 billion marker particles.

\subsection{Comparison between EM and ES turbulence}\label{subsec:results}

\begin{figure}
 \centering
 \includegraphics{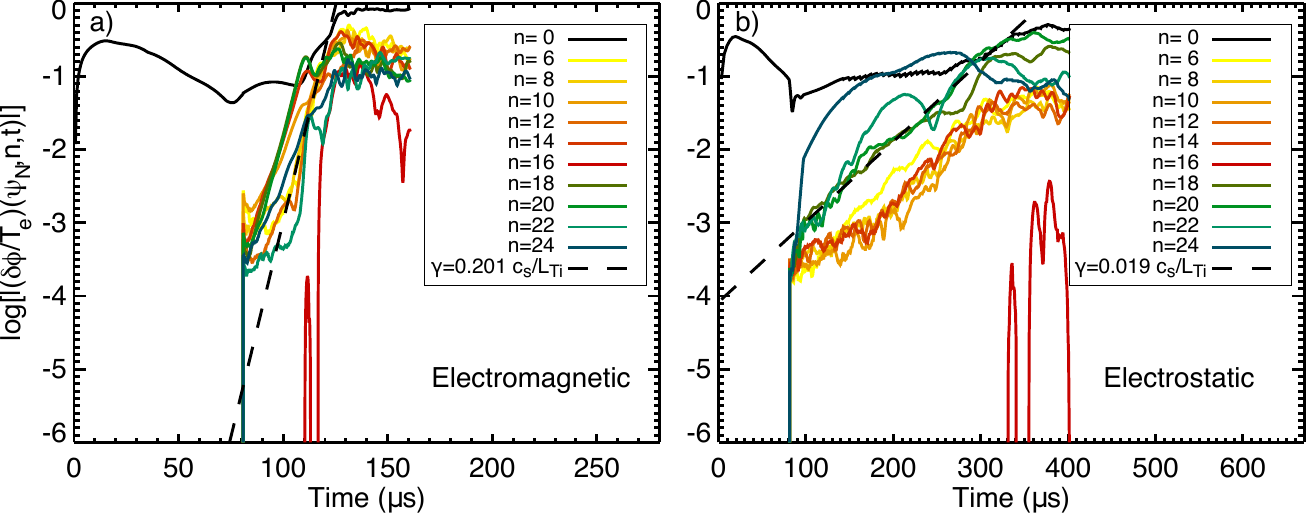}
 \caption{Intensity of the electrostatic potential perturbation $\delta\phi = \phi-\langle \phi \rangle$ for toroidal Fourier modes between $n=0$ and 26 vs. time in the middle of edge pedestal $\psi_N=0.972$ for the (a) electromagnetic and (b) the electrostatic simulation. The dashed line indicates the approximate quasilinear growth rate of the non-axisymmetric mode using the highest amplitude mode as the nonlinear saturation sets in. 
 }
 \label{fig:turb_intensity}
\end{figure}

The intensity of the perturbed electrostatic potential $\delta \phi = \phi- \langle \phi \rangle$ for toroidal mode numbers $n=0,6,8,\dots, 26$ at $\psi_N=0.972$ is shown in Fig. \ref{fig:turb_intensity} (a) for the electromagnetic and (b) for the electrostatic simulation.
The time required to reach nonlinear saturation in the electromagnetic (EM) simulation is much shorter than in the electrostatic (ES) simulation.
In the electromagnetic simulation, the initial growth phase transitions into saturation at $t \gtrsim 120 \,\musec$.
The electrostatic simulation reaches saturation much later at $t \gtrsim 330 \,\musec$.
The ``quasilinear'' growth rate of the dominant toroidal mode in the saturated state is about ten times greater in the electromagnetic ($n=6$, $\gamma\simeq 0.2 c_s/L_{T_i}$ with $c_s=(k_B T_e/m_i)^{1/2}$ and $L_{T_i}$ the ion temperature gradient scale length) than in the electrostatic case ($n=20$, $\gamma\simeq 0.02 c_s/L_{T_i}$).
The toroidal mode spectrum in the electromagnetic study is relatively dense with all modes between $n=6$ and 24 having comparable amplitudes.
In the electrostatic study, the mode numbers $n=18$ and 20 have considerably higher amplitudes than all other modes.
Some of the modes shown in the figures are rather physically irrelevant due to their low amplitude, e.g., $n=16$ in both the electromagnetic and electrostatic simulations.

The long turbulence saturation time in the electrostatic case is somewhat atypical in the H-mode pedestal.
Because of the large gradients in H-mode edge plasma, electrostatic XGC simulations of DIII-D H-modes typically exhibit strong TEM activity with turbulence saturation times of the order of $100\,\musec$ (e.g. Ref. \onlinecite{hager_2020}).
The reason for the low growth rate observed here may be the low toroidal mode number cutoff that is used in the edge pedestal to save computing resources.

To help classify the observed turbulence, we calculate the poloidal-temporal spectrum of the non-axisymmetric component of the electrostatic potential
\begin{equation}\label{eq:dphi_spec_definition}
    \tilde{\delta \phi}(m,\omega) = \left\lbrace \frac{1}{N_\varphi}\, \sum_{i=0}^{N_\varphi-1} \left| \mathcal{F}\left[ \delta\phi \left( \theta^\ast,\varphi=2\pi/i,t \right) \right] \right|^2 \right\rbrace^{1/2},
\end{equation}
where $\mathcal{F}$ indicates the Fourier transform in the straight field-line poloidal angle $\theta^\ast$.
The time intervals used for this calculation are $125\,\musec \leq t \leq 161\,\musec$ in the electromagnetic, and $330 \,\musec \leq t \leq 402 \,\musec$ in the electrostatic case.

\begin{figure}
 \centering
 \includegraphics[width=\textwidth]{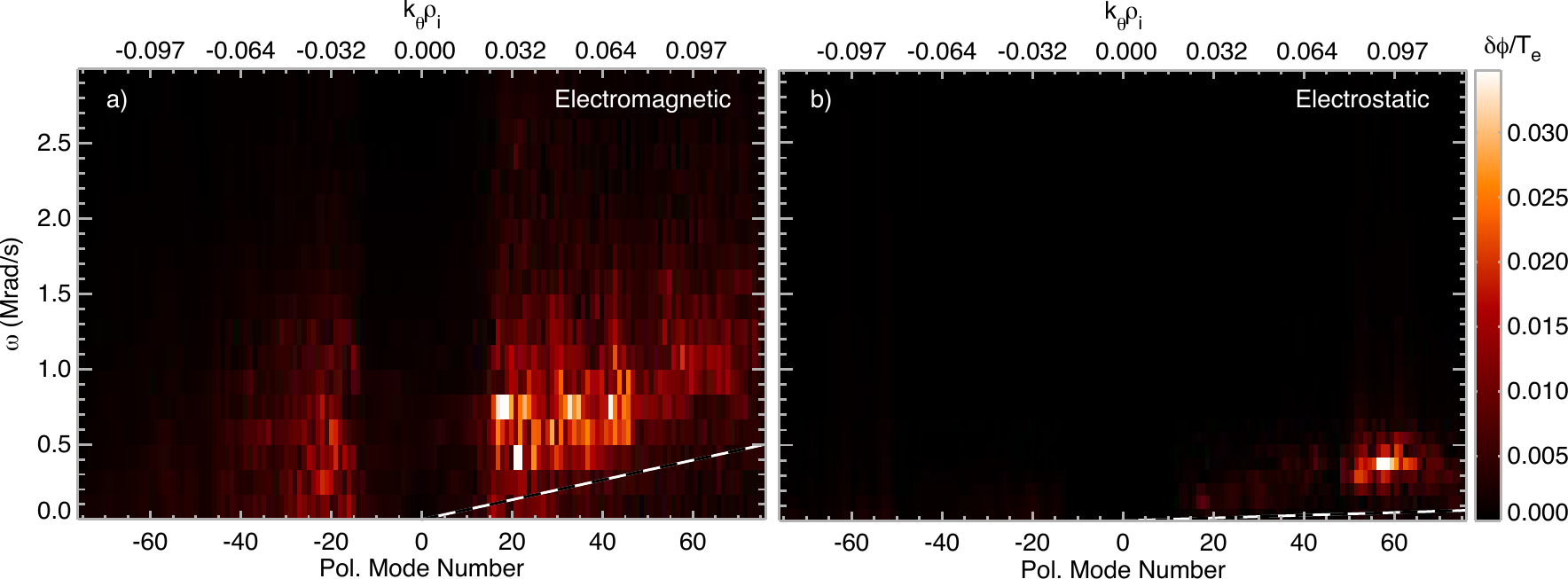}
 \caption{Spectrum of the non-axisymmetric part of the electrostatic potential at $\psi_N=0.972$ from the (a) electromagnetic ($125 \,\musec \leq t \leq 161 \,\musec$) and (b) electrostatic ($330 \,\musec \leq t \leq 402 \,\musec$) simulation.
  The dashed lines indicate the poloidal $\exb$-drift velocity. Positive phase velocity is in the electron diamagnetic direction at the outer midplane.}
 \label{fig:m_om_spectrum}
\end{figure}
\begin{figure}
 \centering
 \includegraphics[width=\textwidth]{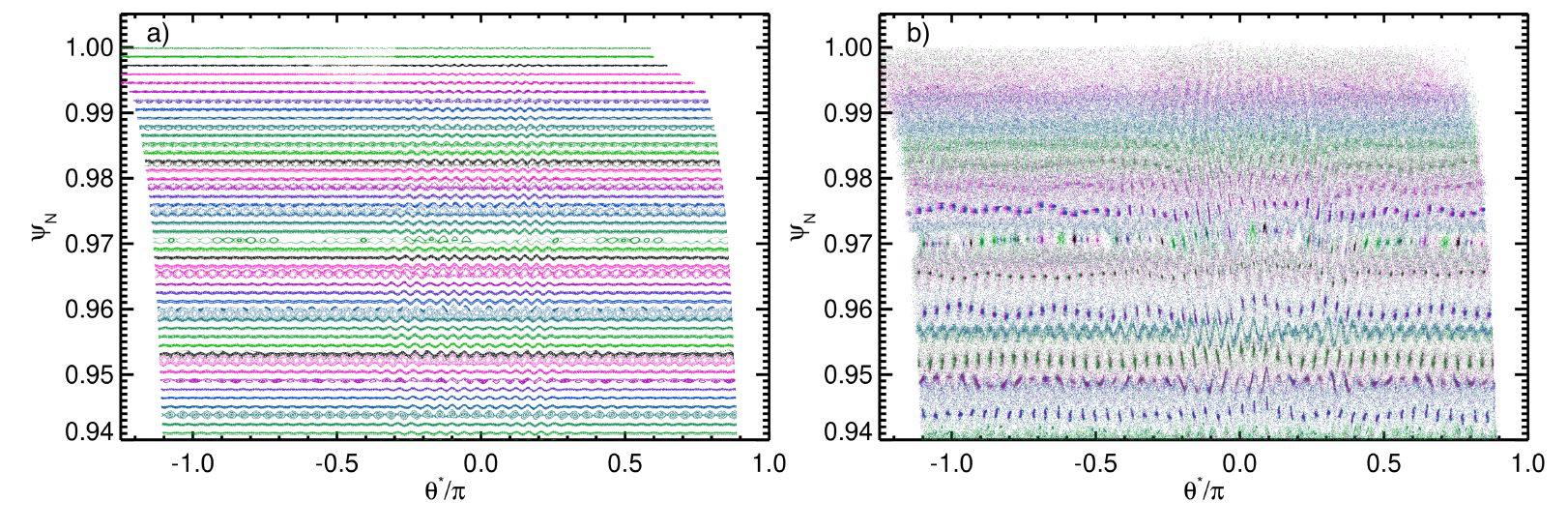}
 \caption{Poincare maps of the perturbed magnetic field at $t=$ (a) 100$\,\musec$ (in the turbulence growth phase), and (b) 160 $\musec$ (after nonlinear saturation in the edge). $\theta^\ast=0$ is the outer midplane, and the minimal and maximal values of $\theta^\star$ are at the inner midplane. Eight perturbed field lines starting at uniformly spaced poloidal angles were traced on each of 112 unperturbed flux-surfaces between $0.85 \leq \psi_N \leq 1$ (total of 896 field lines) for these maps for, on average, 700-800 toroidal circuits.}
 \label{fig:field_lines}
\end{figure}

The resulting spectra at $\psi_N=0.972$ are quite different between the electromagnetic and electrostatic simulations, as shown in Figs. \ref{fig:m_om_spectrum} (a) (EM) and (b) (ES) together with the time and flux-surface averaged $\exb$-velocity.
Positive phase velocity in the spectra corresponds to the electron diamagnetic direction.
Both the electromagnetic and electrostatic simulations exhibit spectral intensity at positive phase velocities faster than the $\exb$-velocity.
For the electrostatic simulation, this implies that the observed modes are trapped electron modes (TEMs).
In the electromagnetic simulation, TEMs and microtearing modes (MTMs) are possible candidates.
The difference between the growth rates observed in the two simulations (Fig. \ref{fig:turb_intensity}), the higher fluctuation frequency, and the island structured magnetic fluctuations in the electromagnetic case (see Fig. \ref{fig:field_lines}) make MTMs the more likely candidate.
The existence of MTMs as the dominant instability at the low values of $k_\theta \rho_i$ in our numerical study would be consistent with earlier numerical findings, e.g. Refs. \onlinecite{doerk_2015,Coury2016}, and is supported by the existence of  island structures in the Poincar\'{e} maps of the perturbed magnetic fields, which are shown in Figs. \ref{fig:field_lines} (a) at $t=100\,\musec$  (growth phase) and (b) at $t=160 \,\musec$ (saturation phase).

The weakness of the TEM instability in the electrostatic simulation is unsurprising due to the low upper bound on $k_\theta \rho_i \lesssim 0.1$ around the magnetic separatrix in our test problem [see Fig.~\ref{fig:initial_profiles}(b)].
TEMs in the steep-gradient region of H-mode edge plasma, are much more unstable at higher $k_\theta \rho_i\gtrsim 0.3$ similar to Ref. \onlinecite{hager_2020}.

\begin{figure}
 \centering
 \includegraphics{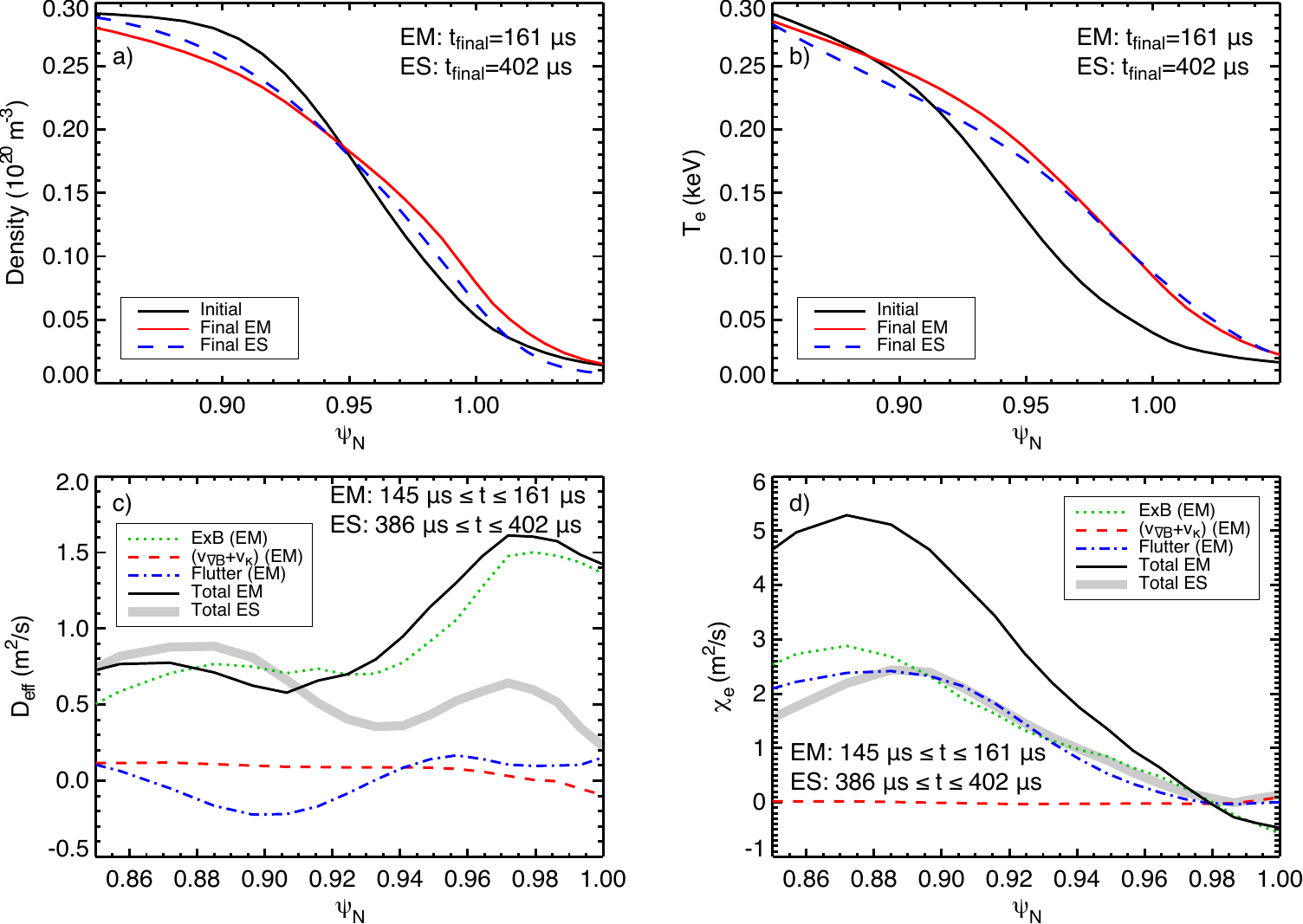}
 \caption{Initial ($t=81 \,\musec$) and final (a) density and (b) electron temperature profiles. Time-averaged effective (c) particle diffusivity $D_{eff}$ and (d) electron heat conductivity $\chi_e$ with contributions from the $\exb$, magnetic inhomogeneity ($v_{\nabla B}+v_\kappa$, ``neoclassical''), and flutter transport channels. In the electrostatic simulation, the flutter transport vanishes, and the magnetic inhomogeneity transport is negligible.
 }
 \label{fig:prof_transp}
\end{figure}

The final flux-surface averaged density and electron temperature are also different between the electromagnetic and electrostatic simulations, as shown in Figs. \ref{fig:prof_transp} (a) and (b).
The final time is $t_{final}^{(ES)}=402 \,\musec$ for the electrostatic simulation, and $t_{final}^{(EM)}=161 \,\musec$ for the electromagnetic simulation.
Since we do not include the higher $k_\theta \rho_i$ fluctuations in the magnetic separatrix area in this demonstration simulation, this difference is not conclusive and requires more accurate study in the future.

We analyze the transport fluxes by splitting the total electron particle and heat flux densities in three components analogous to Ref. \onlinecite{hager_2020} that originate from: i) the $\exb$ drift, ii) the (``neoclassical'') $\nabla B$ and curvature drifts ($v_{\nabla B}+v_\kappa$), and iii) the flutter flux $v_\parallel \delta B_r/B$ from the radial magnetic field perturbation $\delta B_r = \delta \vek{B}\cdot \nabla\psi_N/|\nabla \psi_N|$.
Using these flux-densities, we evaluate the effective particle diffusivity $D_{eff}$ and the effective electron heat conductivity $\chi_e$.
%

The time-averaged (ES: $386 \,\musec \leq t \leq 402 \,\musec$, EM: $145 \,\musec \leq t \leq 161 \,\musec$) particle diffusivity and electron heat conductivity are shown in Figs. \ref{fig:prof_transp} (c) and (d).
From the electromagnetic simulation, we show the total (black solid line) transport coefficients, as well as the $\exb$ (dotted green line), the ``neoclassical'' (red dashed line), and the flutter (blue dashed-dotted line) contributions.
From the electrostatic simulation, only the total is shown because there is no flutter contribution, and the ``neoclassical'' contribution is negligible compared to the $\exb$ contribution.
The electromagnetic effective particle diffusivity is up to about twice as large as the electrostatic one at $\psi_N \gtrsim 0.92$.
The flutter particle transport is a relatively small contribution compared to the $\exb$ transport.
This is consistent with earlier observations that the radial electric fields developing in response to the magnetic perturbation are a strong inhibitor of particle flutter transport \cite{hager_2019_1,Yoo_2021,dchen_2018}.
It is interesting to see that the magnetic drift shows a small tendency for inward particle transport across the magnetic separatrix, but is dwarfed by the outward $\exb$ convective particle transport in the electromagnetic simulation.
Between $0.85 \lesssim \psi_N \lesssim 0.98$, the electromagnetic electron heat conductivity is also about twice as large as the electrostatic one.
But in the case of $\chi_e$, unlike the particle transport, the flutter transport channel is responsible for about half of the total $\chi_e$.
The flutter electron heat flux becomes small at $\psi_N\gtrsim 0.98$ and the total $\chi_e$ becomes weakly negative.
(This may be a transient effect due to the short time window available for averaging.)
Figures \ref{fig:prof_transp} (c) and (d) hint at the importance of electromagnetic turbulence in understanding the electron transport in the edge pedestal and across the magnetic separatrix surface.
The small effective radial $\chi_e$  across the magnetic separatrix should be taken with a grain of salt, though, because the the non-Fickian transport dynamics parallel to $\vek{B}$ might not be accurately captured (due to limitations in the spatio-temporal resolution) in the local evaluation of the transport fluxes 
in the partially stochastic magnetic field across the separatrix.
The non-Fickian transport dynamics across two different magnetic topological regions are the subject of a future report.

\section{Summary and Conclusions}\label{sec:summary}

The reduced $\delta f$ implementation of the mixed-variable/pullback method previously verified in XGC \cite{Cole2021} for electromagnetic turbulence simulation in tokamak core plasma has been extended to a total-$f$ implementation to enable electromagnetic gyrokinetic simulation of tokamak edge plasma across the magnetic separatrix, i.e., from the magnetic axis to the material wall.
For the first time, XGC's unique edge physics self-consistently combine neoclassical and electromagnetic turbulence physics, a fully nonlinear Fokker-Planck-Landau collision operator, heat and torque sources, and Monte-Carlo neutral recycling.
XGC's efficient parallelization and scalability can now be used to study electromagnetic microturbulence effects in realistic diverted tokamak geometry.
Details of the numerical method we use are described in Sec. \ref{sec:algorithm}, including a review of the mixed-variable equations of motion, the total-$f$ enabled pullback transform and weight evolution, the spatio-temporal discretization, the handling of the axisymmetric field components, and the use of field-aligned Fourier filtering to avoid Alfv\'{e}n waves with unnecessarily high frequency.

XGC's new capability is demonstrated using a DIII-D H-mode test plasma with modified density and temperature profiles.
For the purpose of this demonstration, we chose a scenario with low toroidal magnetic field ($B_0\approx 0.7$ T) and used relatively low resolution to save computational resources. 

Comparison between the total-f electrostatic simulation and the total-$f$ electromagnetic simulation in the demonstration set up shows much faster initial turbulence growth in the strong-gradient edge region in the electromagnetic case than in the electrostatic case even though the saturation amplitudes of the dominant toroidal mode numbers are similar.

In both cases, the poloidal phase velocity of the turbulence modes is in the electron diamagnetic direction in the $\exb$-drift frame of reference.
The unstable pedestal modes in the electrostatic case are identified as trapped electron modes, while they are microtearing modes in the electromagnetic modes in the demonstration setup.
The comparison verifies that electromagnetic simulation is necessary for a higher fidelity understanding of the electron particle and heat transport in the edge pedestal.
Due to the modification to the pedestal plasma profile and the removal of the low toroidal mode numbers in order not to provoke the global MHD instabilities, the simulation results may not represent the realistic edge physics of the DIII-D plasma.
More quantitative electromagnetic edge simulation results will follow this report.

%
\section*{Acknowledgments}
The authors acknowledge the pioneering work by M. D. J. Cole and A. Mishchenko in implementing the mixed-variable/pullback method in the simplified $\delta f$ explicit version of XGC \cite{Cole2021}, which allowed the extension of the scheme into total-$f$ as described in the present report.

Funding for this work was provided through the Scientific Discovery through Advanced Computing (SciDAC) program by the U.S. Department of Energy Office of Advanced Scientific Computing Research and the Office of Fusion Energy Sciences under Contract No. DE-AC02-09CH11466 to Princeton University for Princeton Plasma Physics Laboratory.

Computing resources were provided by the Innovative and Novel Computational Impact on Theory and Experiment (INCITE) program, and the Energy Research Computing Allocations Process (ERCAP).
This research used resources of the Oak Ridge Leadership Computing Facility (OLCF, DE-AC05-00OR22725), the Argonne Leadership Computing Facility (ALCF, DE-AC02-06CH11357), and the National Energy Research Scientific Computing Center (NERSC, DE-AC02-05CH11231, award no. FES-ERCAP0020931), which are U.S. Department of Energy Office of Science User Facilities. 

\textbf{Data availability statement:} Data used in preparing this article will be available in the PPPL Theory Department ARK, see Ref. \onlinecite{xgc_data}.

\textbf{Disclaimer:} The publisher, by accepting the article for publication, acknowledges that the United States Government retains a non-exclusive, paid-up, irrevocable, world-wide license to publish or reproduce the published form of this manuscript, or allow others to do so, for United States Government purposes.

\newpage
\renewcommand{\refname}{ }
\bibliographystyle{aipnum4-2}
\bibliography{references}


\end{document}